\documentclass[twocolumn,rvb,showpacs,amsmath,amssymb]{revtex4}

\usepackage{graphicx}

\begin{document}

\newcommand\br[1]{$\langle#1\rangle$}
\newcommand\dst{\displaystyle}

\title{Phase diagram of 3D ANNNI model
in an effective-field approximation}

\author{Anton \v{S}urda}
 
\affiliation{Institute of Physics Slovak Academy of Sciences
D\'ubravsk\'a cesta, 842 28 Bratislava
}

\date{\today}

\begin{abstract}

An effective-field method for calculation of thermodynamic properties of 
three-dimensional  lattice spin models is developed. It is applied to the
ANNNI model on the simple cubic lattice. The phase diagram of the model, consisting of a large number 
commensurate phases and of an incommensurate phase, is calculated,  confirming the results of previous approaches. 
The phase transition lines for a number of commensurate structures are localized and a strong evidence for absence of the direct phase transition between commensurate phases and the  disordered phase is found.
\end{abstract}


\maketitle

\section{Introduction}

In this paper we  study the ANNNI model on a simple cubic or tetragonal
lattice. This model was first introduced by Elliott \cite{ell} in order to understand
modulated magnetic materials. It is reviewed by Selke and Yeomans
\cite{sel1,sel2,yeom}.
The model is known  to form a low temperature ferromagnetic phase for a small
next-nearest-neighbour interaction and a  $\langle2\rangle$  phase for a
large one. The wedge in the nnn interaction-temperature phase diagram between
this two phases is, at low temperatures, filled by infinite number commensurate
phases.

Theoretical study of the ANNNI model has been based on a large number 
of various   approaches.
The devil's staircase structure  of the phase diagram  at low and medium temperatures was elucidated by  
low-temperature  series expansion \cite{fi1} and mean-field approximations 
\cite{boe, bak1, fi2, sel3, jen}. 
The incommensurate phase was also treated by 
the free fermion approximation \cite{vil, ruj}.
Recently, an anisotropic scaling at the Lifshitz point was used to calculate several 
critical exponents  at this point \cite{hen}. 
A considerable effort was also devoted to investigation of ANNNI thin films 
\cite{sel4,sel5,sel6}.

The mean-field approximations describes qualitatively well the phase diagram of the ANNNI model, 
nevertheless, some of its features were challenged by other approaches, e.g.
 the stability of the commensurate phase 
up to the transition line to the disordered phase. 

To improve the performance of mean-field treatment of the ANNNI model, we develop 
an effective field method, 
which is a generalization of the 
cluster transfer-matrix  method successfully applied to 2D spatially modulated structures 
\cite{sur1, kar, paj}.

Our effective field method resembles the nonlinear mapping approach of Bak \cite{bak2,jen}, but, 
instead of magnetization,
it maps a large number of effective fields, which simulate the cluster environment. It is related 
also to the DMRG method \cite{sur2},  and for the 2D ANNNI model they yield similar  results
\cite{kar, gen2}.
Comparing with DMRG approach, our method is much simpler, 
and instead of diagonalization of density matrix and renormalization of transfer matrix 
by matrix multiplication it  requires only calculation of 
square-root of a function of cluster spin configurations and real-number multiplications \cite{sur2}. 
The results
of our method is in general agreement with other approaches, and it removes the artefacts of the previous 
mean-field methods.

In Section II the 3D ANNNI model is specified, and a new effective field approximation is developed. Results of numerical 
calculations and a tool for distingushing between commensurate and incommensurate phases, which 
lead to construction of the phase diagram are presented in Section III.

\section{Model and method}

We shall generalize the 
cluster transfer-matrix method  (an effective-field
approximation) developed and applied to 2D space-modulated structures some time ago \cite{sur1,paj,kar,sur2}. 
 The
development of the 3D method follows the same ideas that were used
in 2D case, however, a  new approximation in the course of
calculation has to be done. For reasons of clarity the
 method is developed for an ANNNI-type model but it can be easily
 reformulated for any 3D model with short-range interactions.

\begin{figure}
\includegraphics[width=0.7\columnwidth,clip=]{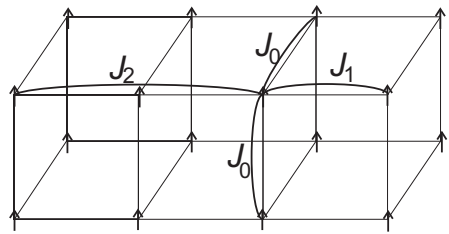}
\caption{\label{mr}
Spin-spin interactions in 3D ANNNI model; nn interactions $J_0$ and $J_1$ are ferromagnetic,
nnn interactions $J_2$ are antiferromagnetic.}
\end{figure}

The three dimensional ANNNI model on a simple cubic lattice consists
of two dimensional planes, within which each spin is coupled to
its nearest neighbors  by a ferromagnetic interaction $J_0$.
However, in the direction perpendicular to the planes, spins are
coupled by competing ferromagnetic nearest-    and
antiferromagnetic next-nearest-neighbor interactions (Fig.~1). For reasons
of simplicity $J_1=J_0$ is further assumed.

As the interactions between spins  $\sigma_{i,j,k}=\pm 1$   in the Hamiltonian
of the 3D ANNNI  model
\begin{eqnarray}
H&=&\sum_{i,j,k} -J_0\sigma _{i,j,k}(\sigma _{i+1,j,k}+\sigma _{i,j+1,k})
-J_1 \sigma _{i,j,k}\sigma _{i,j,k+1}\nonumber\\
&& + J_2 \sigma _{i,j,k}\sigma _{i,j,k+2}
\end{eqnarray}
involve only three layers, it can be written as a sum of layer
Hamiltonians $H_i(S_{i}, S_{i+1},S_{i+2})$ which depend on three
layer variables $S_k\equiv \{\sigma _{i,j,k}\}\ i,j\in(-\infty,
\infty)$.  Since there are only nearest-neighbor  interactions
inside the layers, the layer Hamiltonian can be expressed as a sum
of cluster Hamiltonians defined on $2\times2\times3$ clusters with
the longer side oriented along the  $J_2$ interaction.
\begin{equation}\label{2}
H=\sum_k(H_k(S_k,S_{k+1},S_{k+2})=\sum_k\sum_{i,j}H_{i,j,k}(\sigma _{l,m,n})
\end{equation}
where  $l=i,i+1,\ m=j,j+1,\ n=k,k+1,k+2$.

The exponential of the layer Hamiltonian $H_k$ is further denoted by  $T_k
(S_k, S_{k+1},S_{k+2})\equiv\exp[\beta H_k(S_k,S_{k+1},S_{k+2})]$  and
sometimes
called transfer matrix though it is rather a function of spin variables.

Then the summation in the  partition function
\[
Z=\sum_{\{\sigma _i\}}\exp[\beta H(\sigma _i)]
\]
may be performed consecutively layer by layer generating a set of
auxiliary functions $\Psi_k$ and normalization factors $\lambda_k$
\begin{equation}
\sum_{S_k} \Psi_k(S_k, S_{k+1})T_k(S_k, S_{k+1}, S_{k+2})=\lambda _k
\Psi_{k+1}(S_{k+1}, S_{k+2})  
\end{equation}
starting from an appropriate function $\Psi_1(S_1,S_2)$ that may
be interpreted as a boundary condition of the system on a
semi-infinite lattice.  The values of $\Psi_k$ for  $k\to \infty$
mostly do not  depend on the input $\Psi_1$  except the vicinity
of a first order phase transition.  Here the different bulk values
correspond to one stable and one or more physically unstable
solutions. The stable solution is the one with the lowest free
energy that is proportional to $\log\prod_k \lambda _k$.

As we see, the auxiliary functions in the transfer matrix method
are some general positive functions defined on clusters of planes
in 3D models. For lower-dimensional models they are defined on
clusters of rows in 2D and clusters  of sites in 1D. In a
one-dimensional model, the auxiliary functions depend on finite
number of spin variables, in 2D and 3D cases they acquire infinite
number of values which cannot be generally found by numerical
calculations. Instead of the whole function at the right-hand side
of (3), we further calculate only its  correlation function, sum
of $\Psi_{k+1}(S_{k+1}, S_{k+2})$ over the whole lattice except a
small cluster,  and the true auxiliary function $\Psi$ is
approximated by a more convenient one, nevertheless, exactly
reproducing the calculated correlation functions.

As we do not use any further information from the left-hand side
of Eq. (3), all the remaining properties of the approximate
function $\tilde\Psi$ are derived from the requirement of maximum
of the information entropy $S=\tilde\Psi\log\tilde\Psi$
 \cite{jaynes}. To maximize  $S$
 under the condition
that the partial sum of $\tilde\Psi$ is equal to the given
correlation function, Lagrange multipliers corresponding to each
configuration of the cluster have to be introduced. It is easy to
show that the desired auxiliary function can be expressed as a
product of exponentials of the cluster Lagrange multipliers. Thus,
the requirement of maximum of the information entropy leads to a
factorization of the auxiliary function. If only factorized
auxiliary functions are used, the left-hand side of (3)  is
completely factorized  for short-range interactions  and its
partial summation is equivalent to calculation of a correlation
function of a statistical system of the dimension lower by 1 than
that of the original problem. It means that for 2D system this
step can be performed exactly, but for 3D the factorization
procedure must be applied even for calculation of the correlation
function.

In the case of 2D model (1D auxiliary functions) the  application
of the above considerations  is straightforward. Let us denote the
correlation function on a small cluster of the length $n+1$ by
\[
\dst\Phi_i^{i+n}\equiv\hskip-15pt \sum_{\{\sigma_j\} \atop j\in
(-\infty, i)(i+n, \infty)}\hskip-15pt\Psi(\sigma _j),
\]
 which is
assumingly known from the previous calculation step. The
approximate function $\tilde\Psi$  is defined as a product of
unknown cluster functions $\Theta_i^{i+n}(\sigma _j)$ defined on
$n+1$  sites:
\begin{equation}
\tilde\Psi\equiv \prod_i \Theta_i^{i+n}(\sigma _j)\qquad
j\in\langle i,i+n\rangle.
\end{equation}
We would like to express the cluster functions $\Theta_i^{i+n}$ in terms of
$\Phi_i^{i+n}$.

Let us denote the left eigenfunction of the function (transfer
matrix) $\Theta_i^{i+n}$
\begin{equation}
\sum_{\sigma _i} \theta_i^{i+n-1}(\sigma _j) \Theta_i^{i+n}(\sigma _j)=
\lambda \theta_{i+1}^{i+n}(\sigma _j)  
\end{equation}
by $\theta$  and its eigenvalue by $\lambda$. ($\theta_i^{i+n-1}$
and $\theta_{i+1}^{i+n}$ are identical function defined on
different clusters if we do not expect any space modulation in
this direction.)

Since $\theta_i^{i+n-1}$  is the result of  summation of $\tilde\Psi$
from $-\infty$  to $i$, correlation function
 $\tilde\Phi_i^{i+n}$  corresponding to  $\tilde\Psi$ can be expressed as
\[
\tilde\Phi_i^{i+n}= \theta_i^{i+n-1}\Theta_i^{i+n}
\bar\theta_{i+n}^{i+1},       
\]
where $\bar\theta$  is the right eigenfunction of $\Theta$ defined by
\begin{equation}
\sum_{\sigma _{i+n}}  \Theta_i^{i+n}(\sigma _j)
\bar\theta_{i+n}^{i+1}(\sigma _j)=
\lambda \bar\theta_{i+n-1}^{i}(\sigma _j).                      
\end{equation}
Since we require  $\tilde\Phi_i^{i+n}=\Phi_i^{i+n}$, the unknown cluster
function $\Theta^{i+n}_i$  is
\begin{equation}
\Theta_i^{i+n}= {\Phi_i^{i+n}\over \theta_i^{i+n-1} \bar\theta^{i+1}_{i+n}}.
 \end{equation}
Unfortunately, the eigenfunctions $\theta$ and $\bar\theta$  are implicit
functions of $\Theta$.  On the other hand,  it can be easily shown that
\begin{equation}
\Theta_i^{\prime\, i+n}= \sqrt{\theta_i^{i+n-1}\over
\bar\theta^i_{i+n-1}} \, \Theta_i^{i+n}\,
\sqrt{\bar\theta_{i+n}^{i+1}\over\theta^{i+n}_{i+1}}        
 \end{equation}
have the same eigenvalues as the original cluster functions $\Theta_i^{i+n}$.
Substituting (7) for $\Theta_i^{i+n}$     we get
\begin{equation}
\Theta_i^{\prime\,i+n}=
{\Phi_i^{i+n}\over\sqrt{\theta_i^{i+n-1}\bar\theta_{i+n-1}^{i}
 \theta_{i+1}^{i +n}\bar \theta_{i+n}^{i+1}}}=
{\Phi_i^{i+n}\over\sqrt{\Phi_i^{i+n-1}\Phi_{i+1}^{i+n}}},         
\end{equation}
where  $\Phi_{i+1}^{i+n}=\sum_{\sigma _i}
\Phi_i^{i+n}=\theta_{i+1}^{i +n}\bar \theta_{i+n}^{i+1}$ and
similarly $\Phi_{i}^{i+n-1}$. Thus, obeying the condition of
maximum of  information entropy, relations (4, 9) yield a
possibility to express the approximate chain auxiliary function
$\tilde\Psi$  in terms of the known correlation function
$\Phi_i^{i+n}$
\begin{equation}
\Psi\approx\tilde\Psi=\prod_i \Theta_i^{i+n} = \prod_i \Theta_i^{\prime\,
i+n}
=\prod_i {\Phi_i^{i+n}\over\sqrt{\Phi_i^{i+n-1}\Phi_{i+1}^{i+n}}}
\end{equation}

In the case of 2D auxiliary function (for 3D models) the relation (10) is further
valid, only the indices denote infinite rows
 of  sites rather than  sites. However now, we cannot  expect
that the correlation function $\Phi_i^{i+n}$ defined on $n+1$
infinite rows could be found in previous calculations, but only a
function on a finite cluster of the size $n\times l$. We  denote
it by  $\Phi_{i,j}^{i+n,j+l}$,  where the first indices represent
rows and the second ones  columns of the lattice. The most serious
difference between 3D and 2D models is that the eigenfunction in
(5) cannot be found exactly but it must be factorized, as well.
Then all the functions in the expression for $\Phi_i^{i+n}$ are
factorized similarly as at the left-hand side of Eq. (3), i.e. the
whole procedure that we applied to the  chain auxiliary function
$\Psi$, and that has led to (9), can be applied to $n$-row and
$n-1$-row correlation function $\Phi_i^{i+n}$ and
$\Phi_{i+1}^{i+n}$, $\Phi_i^{i+n-1}$, respectively, appearing in
2D version of (10). We obtain
\begin{equation}
\Phi_i^{i+n}=
\prod_j {\Phi_{i,j}^{i+n,j+l}\over \sqrt{\Phi_{i,j}^{i+n,j+l-1}\Phi_{i,j+1}^{i
+n,j +l}}}\qquad \hbox{etc.}             
\end{equation}
Thus,  by consecutive application of the factorizing procedure
(11) to  all  terms in (10), an approximation  $\tilde\Psi$ to the
function $\Psi$ can be expressed in terms of its cluster
correlation function
\begin{equation}
\Phi_{i,j}^{i+n,j+l}=\hskip-15pt \sum_{{\{\sigma _{km}\}\atop
k\in(-\infty, i)\cup(i+n,\infty)}\atop m\in(-\infty,
j)\cup(j+l,\infty)}\hskip-15pt \Psi(\sigma _{km}). 
\end{equation}
The expression reads
\begin{eqnarray}
&\Psi\approx \tilde\Psi = \prod_{i,j} \Theta_{i,j}^{\prime i+n,j+l}
\hfill\\
&=
\prod_{i,j} {\Phi_{i,j}^{i+n,j+l}
\root 4 \of {\Phi_{i,j}^{i+n-1,j+l-1} \Phi_{i,j+1}^{i+n-1,j+l}
\Phi_{i+1,j}^{i+n,j+l-1} \Phi_{i+1,j+1}^{i+n,j+l} }\over
\sqrt{\Phi_{i,j}^{i+n,j+l-1} \Phi_{i,j+1}^{i+n,j+l} \Phi_{i,j}^{i+n-1,j+l}
\Phi_{i+1,j}^{i+n,j+l} }
}\nonumber                                                           
\end{eqnarray}

Unlike in 1D case, the correlation function calculated from
$\tilde\Psi$  is only approximately equal to that calculated from
$\Psi$. It would be true if we were able to factorize the whole
two-dimensional plane function $\Theta^{\prime i+n}_i$ in 2D
version of (10) and not only each correlation function $\Phi$
separately.

All the functions in (13) are plane dependent in the case of a
modulated structure. Therefore, in the explicit description of the
iteration procedure, the plane index $k$ should be attached to all
correlation and auxiliary function.

The logarithm of the  cluster functions $\Theta'$ may be interpreted as 
effective fields acting on a plane, simulating the effect of the 
  half-lattice already summed up. However, for simplicity,  the functions $\Theta'$
themselves will be called effective fields.

The computational iteration scheme of the cluster transfer-matrix method for
3D ANNNI model is as follows:

1. From the cluster  functions (effective field) $_k\Theta_{i,j}^{\prime i+n,j+l}$
known from the previous step the approximate auxiliary function
$\tilde\Psi_k(S_k,S_{k+1})=\prod_{i,j}
\,_k\Theta _{i,j} ^{\prime i +n,j +l} $ is constructed and
$\Psi_k(S_k,S_{k+1})$  in (3) is replaced by it.

2.  The correlation function $_{k+1}\Phi_{i,j}^{i+n,j+l}$ (12) of
the auxiliary function $\Psi_{k+1}(S_{k+1},S_{k+2})$ is
calculated from (3). As the both functions at the left-hand side
of (3) are factorized this problem is equivalent to calculation of
a correlation function of a 2D lattice model with short-range
interactions  that was discussed above and in previous papers
\cite{sur1,kar,paj} in detail. This task is performed in two steps, and the
approximate factorization utilizing 1D version of (10) is applied
once.

3.  Formula (13) is used and the cluster  functions
$_{k+1}\Theta_{i,j}^{\prime i+n,j+l}$ are found.

4. Calculation is continued for the next plane starting from  the step 1.

It is convenient to take the result of a previous iteration for
the initial condition  of the  calculation at a nearby point in
the parameter space. The bulk values of the cluster
function are obtained after iteration over  few periods of the
commensurate or incommensurate structure. However, the periods of
the commensurate structure sometimes exceed several hundreds of
lattice constants in our calculations. The convergence of the
iteration procedure is very slow near the continuous
incommensurate-disorder phase transition and the steady state were
reached after more than ten thousands steps.

In our actual calculations the length of the cluster edges $n$ and
$l$ was taken equal to 1, i.e. the cluster on which the functions
$\Phi_{i,j}^{i+n,j+l}$ and $\Theta_{i,j}^{\prime i+n,j+l}$ are
defined has 8 sites (elementary cube) and the functions acquire 256
values. Thus, our generalized mean-field approximation utilizes 256
effective  fields instead of one in previous approaches. \cite{jen}

The planes perpendicular to the nnn interaction are ferromagnetic in the ANNNI
model, thus the cluster correlation function and the cluster  function
do not depend on its position in the plane.
$_k\Phi_{i,j}^{i+n,j+l}$ is in fact only  a short-hand notation  of
$\Phi_k({}_kS_{i,j}^{i+n,j+l})$, where $_kS_{i,j}^{i+n,j+l}$ is a spin
configuration of a cluster in the plane $k$.  Similarly,
$_k\Theta_{i,j}^{\prime i+n,j+l} \equiv
\Theta_k^\prime({}_kS_{i,j}^{i+n,j+l})$.

To find the actual structure at the given point of the phase
diagram, it is not necessary to calculate the lattice site
magnetizations.  The structure can be deduced from the plane
dependence of the effective field $\Theta_k^\prime$. In our
approximation it acquires 256 values, but a plot of arbitrary one
of them can be used  to find  the phase diagram. For reason of
simplicity and symmetry, the difference $\psi_k \equiv
\Theta_k^\prime(+)- \Theta_k^\prime(-)$, is plotted where ``+''
and ``$-$''denote  spin configurations of the 8-site cluster with
all the spins up and down, respectively. The sign  of the function
$\psi_k$ is the same as the sign of the magnetization of the
$k$-th row. The ANNNI model structures consist of sequences of
planes   with negative or positive magnetization. As the external
magnetic field is equal to zero, the commensurate structures are
symmetric with respect to spin inversion. Therefore,  only
$|\psi_k|$ is taken into account further. Its periodicity $p$ is
one half of or equal to the structure periodicity if in the
interval $p$ the function $\psi_k$ changes its sign even or odd
times, respectively. A structure consisting repeatedly of $p$
planes with positive magnetization and $p$ planes of negative
magnetization with periodicity $2p$ is usually denoted in
literature as $\langle p\rangle$. More generally, the sequence of
$n$  clusters of the above-mentioned planes of the length $p$
interrupted by one cluster of the length $p-1$ is denoted as
$\langle p^n (p-1)\rangle$.

At high temperature when the convergence is slow and the areas of
commensurate structures are very narrow, or at lower temperatures
when near the accumulation points the periodicity of commensurate
structures tends to infinity, it is often not possible  to perform
the calculation directly at the point of parameter space of
desired properties, because its precise position is not known.
Nevertheless, the structure at it can  be deduced from the
behavior of the effective field in its close vicinity. For this
purpose, we shall further plot $|\psi_{k+p}| - |\psi_k|$ vs.  $
|\psi_k|$, where $p$ is the periodicity  of the function $
|\psi_k|$  somewhere near the point of the parameter space where we
perform the calculation. It is not necessary to plot $|\psi_{k+p}|
-|\psi_k|$ vs. $ |\psi_k|$ for all values of $k$.
 The information, we are interested in, can be found from behavior
of the plot  for the planes  $k_0+np$ $(n=1,2,\dots)$.  $k_0$
should be the number of the  plane closest to a node of the
structure $({\rm sign}\,\psi_{k_0} \ne {\rm sign}\,\psi_{k_0+1})$,
where $|\psi_{k_0}|$ is close to zero and
$|\psi_{k_0}|-|\psi_{k_0-1}|$ is large. Now, $|\psi_{k+p}|
-|\psi_k|=0$ always means that the structure is commensurate with
period $p$ or $2p$ and not that $|\psi_k|$ is close to its maximum
value. The plots will be drawn for $|\psi_k|$ in the range from 0
to its maximum value when a new plane  $k-1$ with a smaller value
of $|\psi_{k-1}|$ appears. Analysis of them will make possible to
distinguish between commensurate and incommensurate structures and
confirm the existence of the accumulation point, where period of
commensurate structures tends to infinity.

\section{Results and discussion}

Results  of our effective-field calculations are
consistent with the phase diagram obtained by the mean-field
approximation and low-temperature expansion [sel1]. However, the
temperatures, at which the phase transitions occur, are more
realistic, and for the exactly soluble case $J_2=0$ the critical
temperature does not deviate more than 1\% from the true value for
the approximation with 256 effective fields.

From our calculation, in accordance with previous results of other
authors, it is possible to conclude  that the phase  diagram
consists of infinitely many commensurate phases which appear
mostly at low temperatures and an incommensurate and disordered
phase at high temperatures.

At low temperature  we have found a ferromagnetic phase, a commensurate
structure with periodicity 4 consisting of a sequence of couples of planes with
alternating magnetization ($\langle 2 \rangle$), a
structure with periodicity  6 ($\langle 3 \rangle$) and combinations of the
last two structures of the type $\langle 2^n3 \rangle$ $n=1,2,3,\dots$.
As the low-temperature region is fairly well described by the low-temperature
expansion,  we concentrate to the medium- and high-temperature properties of
the phases $\langle 4 \rangle$, $\langle 3 \rangle$, $\langle 23 \rangle$ and
the regions in their close vicinity.

\begin{figure}
\includegraphics[width=\columnwidth,clip=]{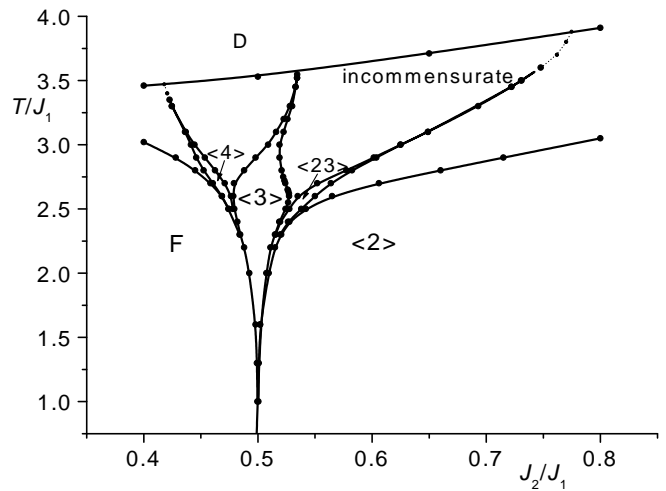}
\caption{\label{2}
Phase diagram of the 3D ANNNI model. Only the basic
structures with short periodicity are depicted. The dotted lines
connect points in the parameter space where the incommensurate
phase has the same periodicity as the corresponding commensurate
structure. The symbols denote the following periodically repeating
structures: \br2 -- $\uparrow\uparrow\downarrow\downarrow$, \br3
-- $\uparrow\uparrow\uparrow\downarrow\downarrow\downarrow$,
\br{23} -- $\uparrow\uparrow\downarrow\downarrow\downarrow$, \br4
-- $\uparrow\uparrow\uparrow\uparrow
\downarrow\downarrow\downarrow\downarrow$, F -- ferromagnetic
$(\uparrow)$, D -- disordered, where the arrows indicate
directions of plane magnetization.}
\end{figure}

The main phases of the 3D ANNNI model obtained from our
calculations are shown in the phase diagram   (Fig.~2). The thick
lines denote the borders of the regions of commensurate phases and
represent first-order phase transition lines. The dotted lines
connect points in the parameter space where the incommensurate
phase has the same periodicity as the corresponding commensurate
structure. The widths of   the commensurate phases  near the
order-disorder phase transition line go to zero for all of them,
i.e., there is no direct transition between the commensurate and
the disordered phase. The commensurate regions at high
temperatures are very narrow (narrower  than the line thickness),
nevertheless, they persist to rather high temperatures. A very
large (probably infinite) number of commensurate phases between
each two main phases are not depicted in the diagram and are
discussed later. The Lifshitz point behind the 
left edge of the diagram is not shown, as the slow
convergence of calculations and complicated phase structure did not 
make possible to correctly interpret the obtained results. 



It is not easy  to prove the existence of the commensurate phase
in a very narrow region and distinguish between the commensurate
and incommensurate phase of the same or a slightly different
periodicity.   Here, it is helpful to observe the above-mentioned
plot of $\Delta\psi\equiv |\psi_{k+p}| - |\psi_k|$ vs.  $|
\psi_k|$, where $p$ is the periodicity of the function $|\psi_k|$
for the assumed commensurate structure.



\begin{figure}
\includegraphics[width=\columnwidth,clip=]{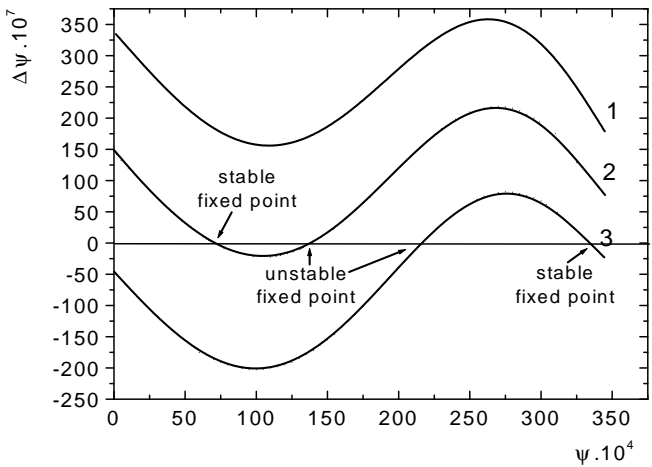}
\caption{\label{3}
Plot of $\Delta \psi \equiv |\psi_{k+3}| - |\psi_k|$ vs.
$\psi\equiv |\psi_k|$ for every third plane of \br3 structure.
$T=3.45$. The plots are drawn for the following values of the
parameters: 1 -- $J_2/J_1= 0.53335$, 2 -- $J_2/J_1= 0.53325$, 3--
$J_2/J_1= 0.53315$. The commensurate \br3 phase is represented by
the stable fixed point. Curve 1 represents an incommensurate
structure. As $\Delta\psi\ll\psi$, the plots are practically
continuous.}
\end{figure}

In Fig. 3  this plot for $T=3.45$ and two different  $J_2/J_1$
inside and near the structure $\langle 3\rangle$
 is shown. $p=3$, and $k$ runs over all planes after which $\psi_k$ changes
its sign, i.e. we plot the function for every third plane. The
structure is commensurate if  $\Delta\psi$  is equal to zero. We
see that for $J_2/J_1= 0.53335$ it never occurs. The function is
incommensurate one with local periodicity greater than 3, and it
changes at different places of the structure, i.e. the true
periodicity is very large, and for decreasing $J_2/J_1$ it tends
to infinity. It can be considered as a phase-modulated  $\langle 3
\rangle$ structure.  As the curve $|\psi_{k+p}| - |\psi_k|$ vs. $|
\psi_k|$ shifts in vertical direction with change of $J_2/J_1$
with only a small change of its shape, we can expect that for some
values of $J_2/J_1$ the curve intersects the $x$-axis and  the
structure becomes commensurate. In Fig.~3 this situation is
exemplified by the curves for   $J_2/J_1= 0.53325$ and 0.53315. In
the bulk the structure is $\langle 3 \rangle$, the difference
$|\psi_{k+p}| - |\psi_k|$ is equal to zero and the structure is
trapped in the stable fixed point. In the transition period, near
the surface or a planar defect, where $\Delta\psi\ne 0$,   the
structure is incommensurate-like. Starting away from an arbitrary
boundary condition the system reaches very fast an incommensurate
metastable state, represented by one of the curves, from which the
stable bulk commensurate structure at the intersection with
$x$-axis slowly develops.

To confirm the existence of the commensurate structure in some
region of parameter space, it is not necessary to find the point
where after many iteration steps the system converges to a bulk
commensurate structure. Near to it $\Delta\psi$  is small and the
convergence is very slow. It is enough to find a nonmonotonous
behaviour of the $\Delta\psi$ vs. $\psi$ plot of an incommensurate
structure somewhere near that parameter space point. A set of such
plots for $T=3.52$ is shown in Fig.~4. It is seen that the
amplitude of modulation of the functions increases when
approaching the commensurate phase. Thus, already a small
modulation of the curve far from the commensurate structure
indicates its presence.

\begin{figure}
\includegraphics[width=\columnwidth,clip=]{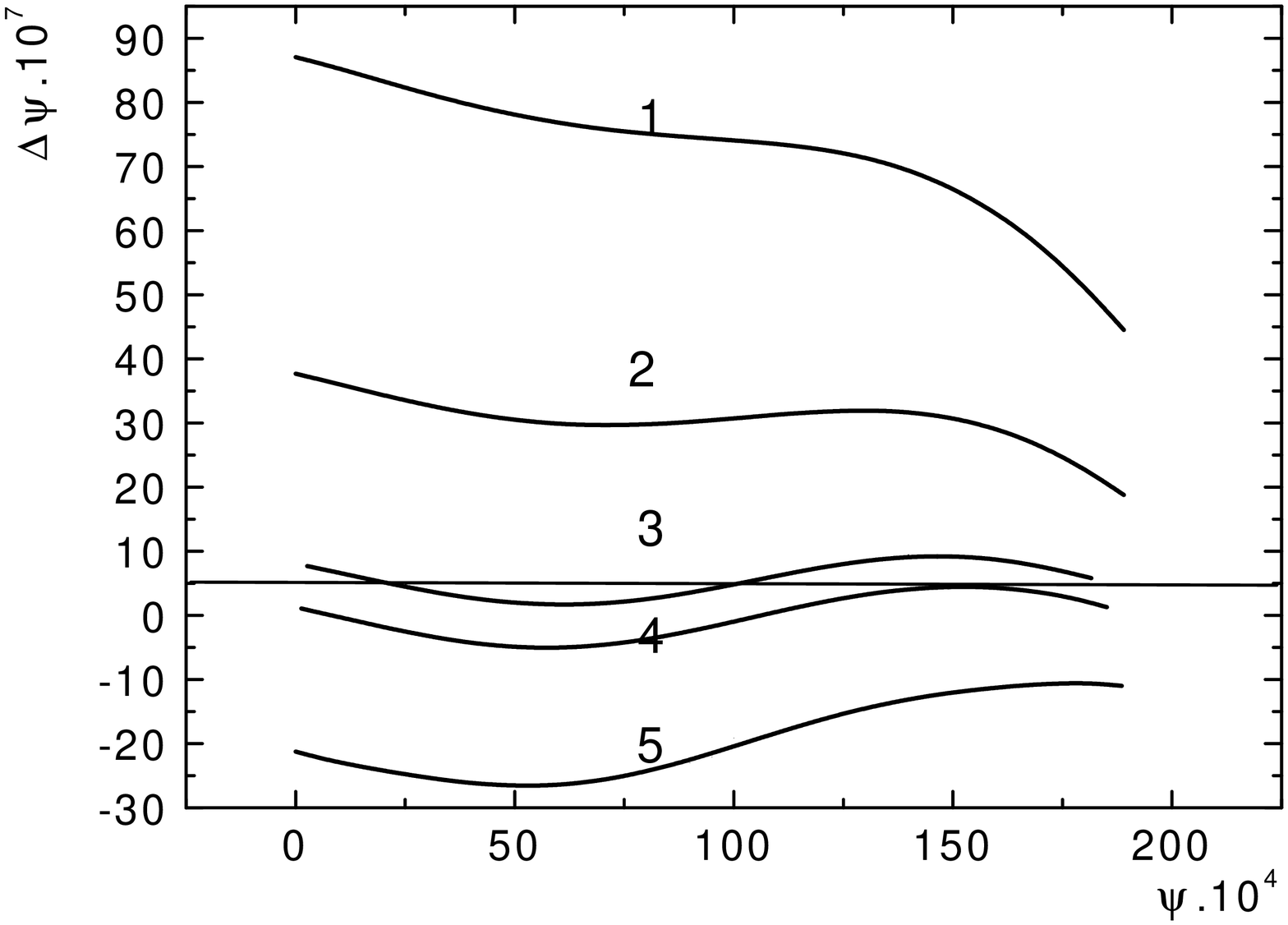}
\caption{\label{4}
Plot of $\Delta \psi \equiv |\psi_{k+3}| - |\psi_k|$ vs.
$\psi\equiv |\psi_k|$ for every third plane and high-temperature,
$T=3.52$, \br3 structure. The plots are drawn for the following
values of the parameters: 1 -- $J_2/J_1= 0.534400$, 2 -- $J_2/J_1=
0.534350$, 3 -- $J_2/J_1= 0.534320$, 4 -- $J_2/J_1= 0.534313$, 5 --
$J_2/J_1= 0.534290$. The commensurate \br3 phase is represented by
curve 4.}
\end{figure}

\begin{figure}
\includegraphics[width=\columnwidth,clip=]{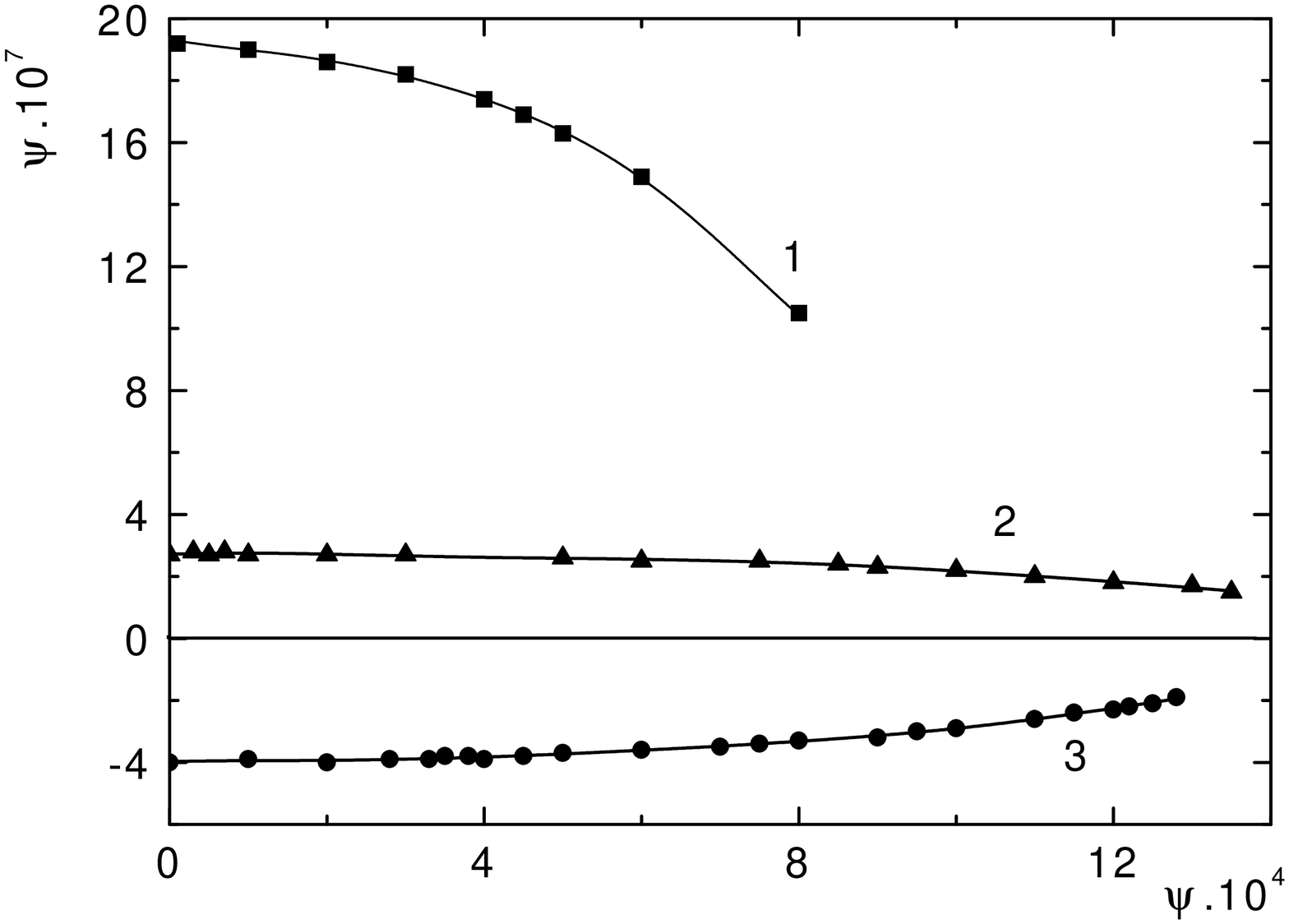}
\caption{\label{5}
In close vicinity of the phase transition line to the
disordered state all the plots
$\Delta \psi \equiv |\psi_{k+3}| - |\psi_k|$ vs. $\psi\equiv |\psi_k|$
are monotonous indicating the absence of commensurate phase in this
region. 
1 -- $T=3.5461,\ J_2/J_1=0.5341$; 
2 -- $T=3.5465,\ J_2/J_1=0.5345$; 
3 -- $T=3.5466,\ J_2/J_1=0.5346$. 
}
\end{figure}

\begin{figure}
\includegraphics[width=\columnwidth,clip=]{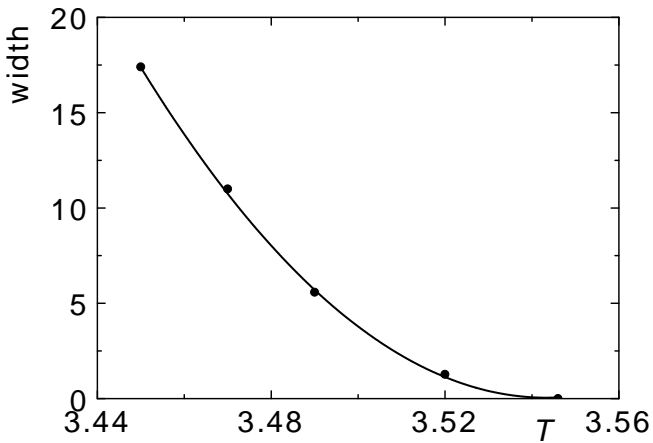}
\caption{\label{6}
Plot of the width of the commensurate phase \br3 vs. temperature. The width is given in the units
of $J_2/J_1\cdot 10^{-4}$.}
\end{figure}

The width of the region of the commensurate structure can be
deduced from the amplitude of the function $|\psi_{k+p}| -
|\psi_k|$ vs. $|\psi_k|$ and the rate of its vertical shift with
change of the parameters.

Fig.~5 shows that for $T=3.546$ no commensurate $\langle3\rangle$
structure exists. All the curves are monotonous. The sign of their
derivatives is negative and positive above and below the $x$-axis,
respectively,  and they do not intersect it. The period of $\psi$
close to 3.    The point in
the parameter space is now very close to the order-disorder phase
transition line, so that the parameters in Fig.~5 should be
carefully changed only in the direction parallel to it. The rate
of convergence is very slow here,  and the bulk incommensurate
structures depicted in the figure were obtained after more than
10,000 iteration steps. The argument of nonexistence of
commensurate structure at $T=3.546$ is confirmed by extrapolation
of the widths of the $\langle3\rangle$ structure to higher
temperatures shown in Fig.~6.  For small values of the width, this
plot could be well fitted by a parabola.

Similar considerations were done for the structures
$\langle4\rangle$, $\langle23\rangle$ and $\langle23^{20}\rangle$
and it was found that the  commensurate structures of higher
periodicity disappear at lower temperatures.  The whole region
near the order-disorder phase transition line is incommensurate
with tongues of commensurate structures of low periodicity which
do not reach the phase transition line.

\begin{figure}
\includegraphics[width=\columnwidth,clip=]{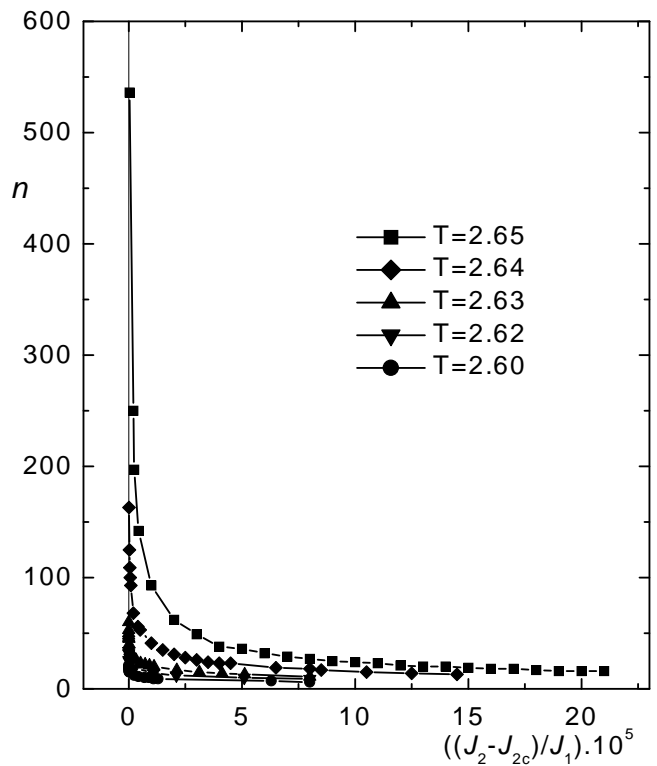}
\caption{\label{7}
Number $n$ of $\uparrow\uparrow\uparrow$ or $\downarrow\downarrow\downarrow$ plane sequences
in \br{23^n} commensurate phases near the transition line to \br3 structure. The value of $J_{\rm 2c}$
at the transition line is different for each temperature. $n$ acquires  discrete values and the
lines are only guides for the eye.}
\end{figure}

At very low  temperatures the phase $\langle 3 \rangle$ is
neighboring to the phase $\langle 23 \rangle$. With increasing
temperature, at $T=1.6$, phases of the type $\langle 23^n \rangle$
start to appear. At given temperature $T$ the period of the
structure,  $p=3n+2$, increases with decreasing $J_2/J_1$ and its
largest value is reached  at the boundary of the $\langle 3 \rangle$
structure. The plots of $n$ near the $\langle 3 \rangle$ boundary
for $T = 2.60, 2.62, 2.63, 2.64, 2.65$ are depicted in Fig.~7. The
periodicity in the close vicinity of $\langle 3 \rangle$ phase
increases very fast and for the temperatures above 2.62 it is not
possible to determine the value of $n_{\rm max}$ or even decide if it is
finite or not. Nevertheless, the accumulation point, where $n_{\rm max}$
becomes infinite, can be found analyzing the plots $|\psi_{k+p}| -
|\psi_k|$ vs.  $|\psi_k|$ for different temperatures and $p=3$,
which are shown in Fig.~8. The structure $\langle23^n\rangle$ for
large $n$ is formed from domains of the structure
$\langle3\rangle$ of the length slightly less than $n$ interrupted
by domain walls symbolically denoted by ``2'' in the symbol
$\langle23^n\rangle$. The $\langle3\rangle$ structure beyond the
wall is shifted by one plane with respect to the structure in the
previous domain. The $\langle3\rangle$ domains correspond to the
minima of the plots in Fig.~8 where the
 $|\psi_{k+3}| -|\psi_k|$ are practically equal to zero. The advent of the wall is
so abrupt and $|\psi_{k+3}| -|\psi_k|$ so large that the next point after the very right
edge of the each
curve is already out of scope of the diagram. With decreasing
$J_2/J_1$ the plots are shifting down and at the $\langle3\rangle$
phase transition line the minimum of the plot touches $x$-axis, and after
some transition period the system remains stuck in $\langle3\rangle$ phase. If the slope of
the plot at minimum is zero, the periodicity near the boundary tends to
infinity, and the
temperature of the system is already above the accumulation point.
From Fig.~8 we see that the
accumulation point is close to the temperature $T=2.64$. The
periodicity of $|\psi|$ tends to infinity if the curve approaches $x$-axis for $J_2\rightarrow J_{\rm2c}$ at $T=2.65$.
Using our method, we were able to find a commensurate structure of
$p=1802$ at this temperature.

\begin{figure}
\includegraphics[width=\columnwidth,clip=]{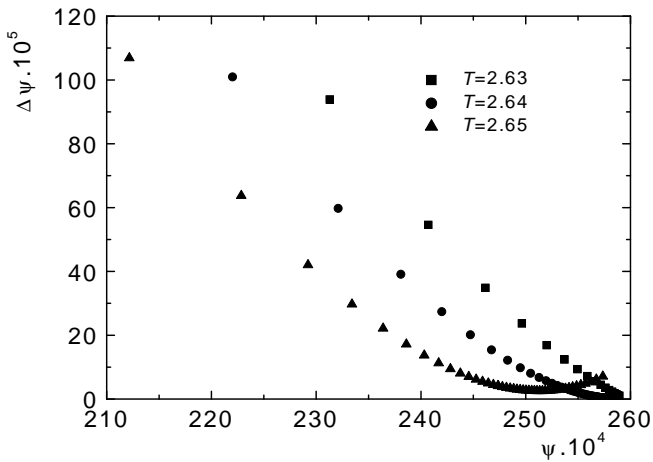}
\caption{\label{8}
Plot of $\Delta \psi \equiv |\psi_{k+3}| - |\psi_k|$ vs. $\psi\equiv |\psi_k|$
for commensurate \br{23^n} structures  near transition to  \br3 phase. $n$ is
equal to the number of points along each curve. (Only points for small value of $\Delta\psi$ are
depicted in the figure.) The curves for $J_{2}=J_{2c}$ touch the $x$-axis. For curves with
zero derivative in the minimum $n$ tends to infinity.}
\end{figure}

For temperatures lower than the temperature of accumulation point  near
the boundary of the $\langle3\rangle$ phase only plain
$\langle23^n\rangle$ phases exist. This is in contradiction with the simple
mean
field approximation findings \cite{sel1}. There is also a difference in location of
the accumulation point. In \cite{sel3} it was found well below the turning point of
the $\langle3\rangle$    boundary where the width of the
$\langle3\rangle$ phase  starts becoming narrower. Our approach locates the
accumulation point $T=2.64$  slightly above   the turning point. Approximately
at the same
 temperature  first combined phases of the type
$\langle (23^n)(23^{n+1}))\rangle$ between $\langle (23^n)\rangle$
and $\langle(23^{n+1}))\rangle$ for large $n$ start to appear.
Similarly to the previous more simple case, following combinations
in the hierarchy are of the type $\langle
((23^n))^k(23^{n+1}))\rangle$ or $\langle
(23^n)((23^{n+1}))^k)\rangle$, which appear at temperatures by 0.04
higher than the temperature of the accumulation point. A great
computational effort is needed to detect a next type of
combinations $(\langle ((23^n))^k(23^{n+1})))^l(\langle
((23^n))^{k+1}(23^{n+1})))\rangle$. These high-order combinations
occupy  very small areas of the parameter space, and with
increasing temperature they are soon replaced by incommensurate
structures.

It is widely believed that the structures with large distances between the
domain walls are commensurate whereas the structures where the distance
between the walls is shorter than the wall-wall interaction are
incommensurate. This statement should be formulated more precisely. The
commensurate  structures with short distances between the walls are more
stable than those with longer distances. They persist to higher
temperatures and they occupy a wider area in the parameter space. Nevertheless,
in the areas between $\langle (23^n)\rangle$ and $\langle(23^{n+1}))\rangle$
for small $n$ the onset of incommensurate structures  was found at lower
temperatures  than for large $n$. The distance between the commensurate
structures increases with decreasing $n$ faster than their width so that there
is enough space for incommensurate structures.

In summary, we developed an effective field approximation, wihich yields by simple iteration procedure practically 
any of probably an infinite number of phases in the phase diagram of the 3D ANNNI model. 
In fact, the method treats  an infinite lattice. 
Lattice size, in contrast to other mean-field and DMRG approaches, does not enter the calculation. 
A difficult task to distinguish between commensurate and incommensurate structure after
a finite number 
of iteration was made easier by plotting the derivative of an effective field
 with respect 
to its value in course if iteration. Our calculations confirmed the general picture of the
phase diagram obtained by other methods, made it more accurate and supported the suggestion 
following from the Monte Carlo calculations that the commensurate phases are separated  from the 
disordered phase by an incommensurate region.

\begin{acknowledgments}
The support by Grant VEGA 2/7174/20 is acknowledged.
\end{acknowledgments}

\end{document}